PAPER • OPEN ACCESS

# Modeling Data Containing Outliers using ARIMA Additive Outlier (ARIMA-AO)

View the article online for updates and enhancements.

# Modeling Data Containing Outliers using ARIMA Additive Outlier (ARIMA-AO)

**Ansari Saleh Ahmar**[1,2*], **Suryo Guritno**[3], **Abdurakhman**[3], **Abdul Rahman**[4], **Awi**[4], **Alimuddin**[4], **Ilham Minggi**[4], **M Arif Tiro**[1], **M Kasim Aidid**[1], **Suwardi Annas**[1], **Dian Utami Sutiksno**[5], **Dewi S Ahmar**[2], **Kurniawan H Ahmar**[2], **A Abqary Ahmar**[2], **Ahmad Zaki**[4], **Dahlan Abdullah**[6], **Robbi Rahim**[7], **Heri Nurdiyanto**[8], **Rahmat Hidayat**[8], **Darmawan Napitupulu**[9], **Janner Simarmata**[10], **Nuning Kurniasih**[11], **Leon Andretti Abdillah**[12], **Andri Pranolo**[13], **Haviluddin**[14], **Wahyudin Albra**[15], **A Nurani M Arifin**[2]

[1]Department of Statistics, Universitas Negeri Makassar, Indonesia
[2]AHMAR Institute, Makassar, Indonesia
[3]Department of Mathematics, Universitas Gadjah Mada, Indonesia
[4]Department of Mathematics, Universitas Negeri Makassar, Indonesia
[5]Department of Business Administration, Politeknik Negeri Ambon, Indonesia
[6]Universitas Malikussaleh, Aceh, Indonesia
[7]School of Computer and Communication Engineering, Universiti Malaysia Perlis, Malaysia
[8]STMIK Dharma Wacana, Lampung, Indonesia
9Lembaga Ilmu Pengetahuan Indonesia, Banten, Indonesia
[10]Universitas Negeri Medan, Medan, Indonesia
[11]Universitas Padjadjaran, Bandung, Indonesia
[12]Universitas Bina Darma, Palembang, Indonesia
[13]Department of Informatics, Universitas Ahmad Dahlan, Yogyakarta, Indonesia
[14]Universitas Mulawarman, Samarinda, Indonesia
[15]Faculty of Economics and Bussiness, Universitas Malikussaleh, Aceh, Indonesia

*ansarisaleh@unm.ac.id

**Abstract**. The aim this study is discussed on the detection and correction of data containing the additive outlier (AO) on the model ARIMA (p, d, q). The process of detection and correction of data using an iterative procedure popularized by Box, Jenkins, and Reinsel (1994). By using this method we obtained an ARIMA models were fit to the data containing AO, this model is added to the original model of ARIMA coefficients obtained from the iteration process using regression methods. In the simulation data is obtained that the data contained AO initial models are ARIMA (2,0,0) with MSE = 36,780, after the detection and correction of data obtained by the iteration of the model ARIMA (2,0,0) with the coefficients obtained from the regression $Z_t = 0,106 + 0,204 Z_{t-1} + 0,401 Z_{t-2} - 329 X_1(t) + 115 X_2(t) + 35,9 X_3(t)$ and MSE = 19,365. This shows that there is an improvement of forecasting error rate data.

**1. Introduction**

In time series data is sometimes there's a far different data values from other data and do not reflect the characteristics of a set of data. The data value is called the outliers. In the analysis of time series data are often obtained outlier. Outlier data is a major impact on forecasting data if we use forecasting methods such as ARIMA, ARMA, and others. Forecasting is an activity to predict what will happen in the future [1]. To forecast a time series then we can use forecasting method, e.g., ARIMA Box-Jenkins method [2]. Forecasting will be far off from what we predict if the data used to predict the data contained outliers. The study on outlier detection is very important because the presence of outliers can lead to parameter estimation or forecasting becomes inappropriate. If a data outlier is not enforced properly, then it will impact the forecast does not reflect the actual data. Related to the incident such as recording and typing errors, AO is the most common type of outliers are found in the time series, so that according to the discussion above, this research will discuss the detection and correction Data containing Additive outlier (AO) in the ARIMA(p,d,q) model.

If the observations in a time series can be predicted with certainty and do not require further investigation, it is called deterministic time series and if the observations can only show the structure of the probabilistic state will come a time series, the time series is called stochastic. In the modeling of time series analysis assumes that the data are stationary. Stationary time series is said if there is no change in the average trends and change in variance. Relatively stationary time series are not increasing or decrease in the value of the extreme fluctuations in the data, or the data is at about the average value of the constant.

Stationarity can be seen by using a time series chart is a scatterplot between the value of the variable $Z_t$ with time *t*. If the diagram time series fluctuates around a line parallel to the time axis (*t*), then the series is said stationary in average. When stationary conditions in the average unmet need differentiation or differencing process [3].

The process of first-order differencing on the difference between tth data with *t*–1 th, i.e., $\Delta Z_t = Z_t - Z_{t-1}$. Forms for second order is differencing $\Delta^2 Z_t = \Delta Z_t - \Delta Z_{t-1} = (Z_t - Z_{t-1}) - (Z_{t-1} - Z_{t-2}) = Z_t - 2Z_{t-1} + Z_{t-2}$. If these conditions are not met stationary in variance, Box and Cox (1964) introduced the power transformation $Z_t^{(\lambda)} = \frac{Z_t^{(\lambda)} - 1}{\lambda}$, ($\lambda$ is a parameter), known as the Box-CoX transformation. Following are some provisions to stabilize the variance is [4]:

- The transformation may be made only for the series $Z_t$ are positive.
- The transformation is done before the differencing and modeling of time series.
- The value selected based on the sum of squares error (SSE) of the transformed series. The smallest value of SSE most constant yield variance.
- Transformation is not only stabilized the variance, but it can also normalize the distribution.

Integrate models Autoregressive Moving Average (ARIMA) has been studied by George Box and Gwilym Jenkins in 1976 [5], and their names are then frequently synonymous with ARIMA processes applied to the analysis of time series. In general ARIMA models is denoted by the notation ARIMA (p, d, q), where p denote the order of the autoregressive (AR), d expressed differentiation (differencing), and q express order of the moving average (MA). Autoregressive model (AR) was first introduced by Yule on 1926 [6] and later developed by Walker on 1931 [7], while model Moving Average (MA) was first used by Slutzky on 1937 [8]. And 1938, Wold generate the theoretical basis of the combination of ARMA and combinations is then often used [9]. Box and Jenkins have effectively managed to reach an agreement on the relevant information needed to understand and use ARIMA models for time series of the variables [10].

In data time series forecasting methods ARIMA (p, d, q) there is the so-called measures or its phases. The stages in the forecasting [11]:

   **a. Identification Model**
   Identification of the model was done to see the significance and stationary of autocorrelation data, so whether or not to do transformation or differencing process (differentiation). From this stage, the model is obtained while testing the model will be made whether or not the corresponding data.
   **b. Assessment and Testing Models**
   After identifying the model, the next step is the assessment and testing of the model. At this stage, divided into two parts, namely parameter estimation and diagnostic model.
   **c. Parameter Estimation**
   After obtaining one or more models while the next step is to find estimates for the parameters in the model.
   **d. Diagnostic Model**
   Diagnostic checking to do check whether the model estimated sufficient or adequate fit with the data. Diagnostic checking is based on the analysis of residuals. The basic assumption is that the residuals of ARIMA models are independent normally distributed random variables with zero mean constant variance.

The outlier is a problem often encountered in financial data. With the data outliers are the parties to a prediction or forecast of the data will have problems such as the problem of lack of accurate in forecasting. Gounder, et al [12] describes several types of outliers are: (1) Additive outlier (AO). Outliers of the simplest and most frequently studied in the analysis of time series are an additive outlier, also known as Type I outliers. An AO only affects a single observation, which could be worth a small or larger value compared with the expected value in the data. After this effect, the data series back to the normal track as if nothing happened. Effects of an AO is independent of ARIMA models and constrained. Outlier AO is the most problematic because it contains two consecutive residues, one before and the other after AO. An AO can have serious effects on the observed properties. This will affect the residual suspicion and supposition parameters. This can be proved in general that the AO large would encourage all autocorrelation coefficients towards zero. The influence of outliers is reduced to a large sample size.

The formula of the AO are as follows [11]:

$$Y_t = \begin{cases} Z_t, & t \neq T \\ Z_t + \omega, & t = T \end{cases}$$
$$= Z_t + \omega I_t^{(T)} \qquad (1)$$

where:
$Y_t$ = Series observations
$Z_t$ = array outlier-free observations
$\omega$ = magnitude outlier
$I_t^{(T)}$ = Indicator variable that indicates the time outlier

(2) innovational outlier (IO), in contrast to AO, an innovational outlier is also known as a Type II outliers (Fox, 1972) that affect some observations. An AO only affects only one residue, at the time of the occurrence of outliers. Effect of IO on the observed series consists of an initial shock that spread in subsequent observations by the weight of the moving average representation (MA) of the ARIMA models. So far the weight is often explosive, may influence the IO, in some cases, the incidence has increased and continues to increase to more and more value from that time until the events of the past, as well as in the future more unwelcome. Effect of IO depends on the particular model of the series, and series with stationary transformation effect indefinitely. For series containing trend and seasonal, the IO will affect both. IO presents some serious drawbacks that should be avoided. The problem then is that the other three types of outliers that can not explain the changes in the seasonal component.

## 2. Method

The method used in this research is the study of literature by analyzing the detection and correction of data AO on the ARIMA(p,d q) model. The initial step in the analysis is to stage in forecasting ARIMA (p,d,q) in order to obtain the corresponding allegations of ARIMA model. Further, testing of stationary residual. If the residual of stationary is not obtained then alleged that occurs because of the data that it contains outliers. Then do the detection and correction of data containing AO with an iterative procedure to obtain the residual of stationary and has the smallest MSE.

## 3. Result and Discussion

The outlier is a problem often encountered in financial data. With the data outliers are the parties to a prediction or forecast of the data will have problems such as the problem of lack of accurate in forecasting.

For a stationary process, say $Z_t$ as the observed series and $X_t$ as an outlier-free series. Assume that $X_t$ follows the general model ARMA (p, q):

$$\phi_p(B)Z_t = \theta_q(B)a_t, a_t \sim WN(0,\sigma^2), \theta_i, \phi_i \in R, t \in R \tag{2}$$

with $\phi_p(B) = 1 - \phi_1 B - ... - \phi_p B^p$ dan $\phi_p(B) = 1 - \phi_1 B - ... - \phi_p B^p$ is a stationary and invertible operators who do not share the same factor. Additive outliers (AO) is defined as:

$$Y_t = \begin{cases} Z_t, & t \neq T \\ Z_t + \omega, & t = T \end{cases} \tag{3}$$

$$Y_t = Z_t + \omega I_t^{(T)} \tag{4}$$

$$Y_t = \frac{\theta(B)}{\phi(B)}a_t + \omega I_t^{(T)}, a_t \sim WN(0,\sigma^2), \theta_i, \phi_i \in R, t \in R \tag{5}$$

where:

$$I_t^{(T)} = \begin{cases} 1, & t = T, \\ 0, & t \neq T, \end{cases} \tag{6}$$

is an indicator variable representing whether or not an outlier at time T.
In general, if the time series data are k additive outlier (AO), the general model of an outlier this:

$$Y_t = \sum_{j=1}^{k} \omega_j I_t^{(T_j)} + \frac{\theta(B)}{\phi(B)}a_t, a_t \sim WN(0,\sigma^2), \omega_j, \theta_i, \phi_i \in R, t \in R \tag{7}$$

### 3.1. Estimated Effect Outlier when Outlier Time Unknown

As a matter of motivation to detect AO procedure, we consider the simple case when T and all the parameters in (3) known. For example:

$$\pi(B) = \frac{\phi(B)}{\theta(B)} = (1 - \pi_1 B - \pi_2 B^2 - ...) \tag{8}$$

and defined:

$$e_t = \pi(B)Y_t \tag{9}$$

from (6) can be established:

$$AO: \quad e_t = \omega \pi(B) I_t^{(T)} + a_t, a_t \sim WN(0, \sigma^2), \omega_i, \pi_i \in R, t \in R \tag{10}$$

For n observations made, AO models above can be written as:

$$\begin{bmatrix} e_1 \\ \vdots \\ e_{T-1} \\ e_T \\ e_{T+1} \\ e_{T+2} \\ \vdots \\ e_n \end{bmatrix} = \omega \begin{bmatrix} 0 \\ \vdots \\ 0 \\ 1 \\ -\pi_1 \\ -\pi_2 \\ \vdots \\ -\pi_{n-T} \end{bmatrix} + \begin{bmatrix} a_1 \\ \vdots \\ a_{T-1} \\ a_T \\ a_{T+1} \\ a_{T+2} \\ \vdots \\ a_n \end{bmatrix} \tag{11}$$

Let $\hat{\omega}_{AT}$ as the least squares estimator of to model AO. Because it is white noise, then the theory of least squares, found that:

$$AO: \quad \hat{\omega}_{AT} = \frac{e_T - \sum_{j=1}^{T} \pi_j e_{T+j}}{\sum_{j=0}^{n-T} \pi_j^2} \tag{12}$$

$$= \frac{\pi^*(F) e_T}{\tau^2}$$

with $\pi^*(F) = (1 - \pi_1 F - \pi_2 F^2 - \ldots - \pi_{n-T} F^{n-T})$, F is the forward shift operator so that $Fe_t = e_{t+1}$ dan $\tau^2 = \sum_{j=0}^{n-T} \pi_j^2$. The variance of the estimator is:

$$Var(\hat{\omega}_{AT}) = Var\left(\frac{\pi^*(F) e_t}{\tau^2}\right)$$

$$= \frac{1}{\tau^4} Var[\pi^*(F) e_t] \tag{13}$$

$$= \frac{\sigma_a^2}{\tau^2}$$

### 3.2. Procedure Using Iterative Outlier Detection

If T is not known, but the parameters of time series are known, then we can calculate $\lambda_{1,t}$ for each $t = 1, 2, \ldots, n$, and then make a decision based on the results of these sampling. However, in practice often be found that parameter time series $\phi_j, \theta_j, \pi_j$ and $\sigma_a^2$ usually unknown and must be predictable. With the outlier then make predictions parameter would be biased. In particular, $\sigma_a^2$ will tend to be exaggerated, as indicated earlier. Therefore it is the Chang and Tiao (1983) proposed an iterative procedure to detect and deal with the situation when the number of AO is not known to exist. In an outlier detection and correction of the data it needs is called procedure. The procedure to be used in the detection procedure is iterations. The procedure is as follows (Box, Jenkins, and Reinsel, 1994):

Step 1: Download the data
Step 2: Estimated parameters of ARIMA (usually assumed that the data contains no outliers)
Step 3: Calculate the residuals of the ARIMA models are:

$$\hat{e}_t = \frac{\hat{\phi}(B)}{\hat{\theta}(B)} Y_t = \hat{\pi}(B) Y_t \tag{14}$$

with :

$$\phi_p(B) = 1 - \phi_1 B - ... - \phi_p B^p \text{ and } \phi_p(B) = 1 - \phi_1 B - ... - \phi_p B^p \tag{15}$$

and let :

$$\sigma_a^2 = \frac{1}{n} \sum_{t=1}^{n} \hat{e}_t^2 \tag{16}$$

Step 4: Calculate the test statistic presence of AO ($\hat{\lambda}_{A,T}$)

$$\hat{\lambda}_{1,T} = \frac{\tau \hat{\omega}_{AT}}{\hat{\sigma}_a} \tag{17}$$

with :

$$\hat{\omega}_{AT} = \frac{e_T - \sum_{i=1}^{T} \pi_i e_{T+i}}{\sum_{i=o}^{n-T} \pi_i^2} \tag{18}$$

$$\tau^2 = \sum_{i=0}^{n-T} \pi_i^2 \tag{19}$$

Jika $|\lambda_{1,T}| > c$ then there is AO on the time $t$.

c is the critical value that is calculated using the formula presented by Ljung in the SAS / ETS User's Guide.

AO effects may remove of residual by defining:

$$\hat{e}_t = \hat{e}_t - \hat{\omega}_{AT} \hat{\pi}(B) I_t^{(T)} = \hat{e}_t - \hat{\omega}_{AT} \hat{\pi}_{t-T}, t \geq T \tag{20}$$

In other cases, new estimates $\hat{\sigma}_a^2$ calculated from the modified residual $\hat{e}_t$.

If there are any outliers are identified the modified residual $\hat{e}_t$ and modification estimates $\hat{\sigma}_a^2$, but with the same parameters $\hat{\pi}(B) = \frac{\hat{\phi}(B)}{\hat{\theta}(B)}$ used to calculate the static of $\hat{\lambda}_{1,T}$. Steps are repeated until all outliers are identified.

Let outlier identification procedure at k point in time, $T_1, T_2, ..., T_k$. So the overall outlier correction model of it:

$$Y_t = \sum_{j=1}^{k} \omega_j I_t^{(T_j)} + \frac{\theta(B)}{\phi(B)} a_t \tag{21}$$

The formula above is estimated for the observation sequence $Z_t$ for AO.
And it with a revised set of residuals:

$$\hat{e}_t = \frac{\hat{\phi}(B)}{\hat{\theta}(B)} \left[ Y_t - \sum_{j=1}^{k} \hat{\omega}_j I_t^{(T_j)} \right] \tag{22}$$

and obtained from this model fit.
To modify the observation time series containing the effects of outliers can remove and correction of the data done by constructing the generalized estimating equations as follows:

$$\hat{z}_t = Z_t + \sum_{j=1}^{k} \hat{\omega}_j I_t^{(T_j)} \tag{23}$$

**4. Simulation**
The data used in this simulation is the data obtained from the ARIMA simulation results using the R application. The steps taken in the simulation forecast above is as follows:

*4.1. Phase Identification*
The first step in the modeling of time series on the identification is checked stationary data by plotting the data as shown below.

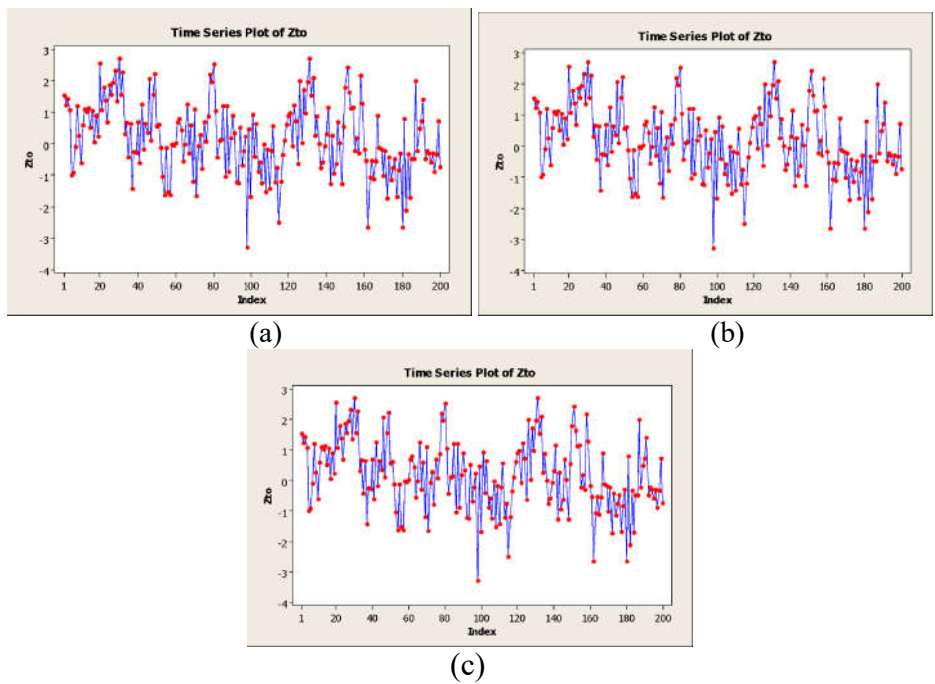

**Figure 1.** (a) Plot Time Series Data, (b) ACF Plot Time Series Data, (c) PACF Plot Time Series Data

From the picture looks stationary in the variance of the data and then do the ACF and PACF plots her. Based on the data diagram, the value of ACF and PACF, the data has been stationary in the average. Once the data is stationary, then the next step is to identify the model temporarily. Identification of the model done by looking at the values of the coefficient autocorrelation, the ACF and PACF plots of the data that has been stationary. From the form of the ACF and PACF are truncated form after a lag of 2 and dropped exponentially suspected that the appropriate model for these data is ARIMA (2,0,0) or AR(2) with parameters $\phi_1 = 0,2237$ and $\phi_2 = 0,4282$.

*4.2. Assessment and diagnostic checks*

The early stages of interpreting the results of time series analysis are to look at its significance model parameters that have been modeled. The model is the first approximation obtained AR(2), so we need to test the parameters of Autoregressive $(\phi)$.

Estimated parameters of the model AR(2) is significantly different from zero with 95% confidence level. This can be seen in the p-values. For parameter AR(1) that $\phi_1$, p-value = 0.001 and for AR(2) is $\phi_1$, p-value = 0.000. If the calculations are performed with a computer statistical package, then simply use the p-value already known. Criteria conclusion that significance testing if p < α and not significant when p ≥ α. Based on the results of Minitab for the data obtained p-value = 0.001 < α = 0.05, which means significant testing. After testing its significance parameters, the next step is to test the suitability of the model. To suitability models include the adequacy of the model (test whether the remaining white noise) and test the assumption of a normal distribution. In Minitab results also can be seen that the value of Ljung-Box statistics is displayed on lag 12, lag 24 and lag 36. Ljung-Box statistic value at lag 12, lag 24 and lag 36 consecutive shows p-value = 0.619, 0.520, 0.203, and 0.177 p-value is greater than α = 0.05. This means that the remaining eligible white noise.

Next, test the remaining normal distribution. The results of the test chart normal distribution for the residual shown in figure 2.

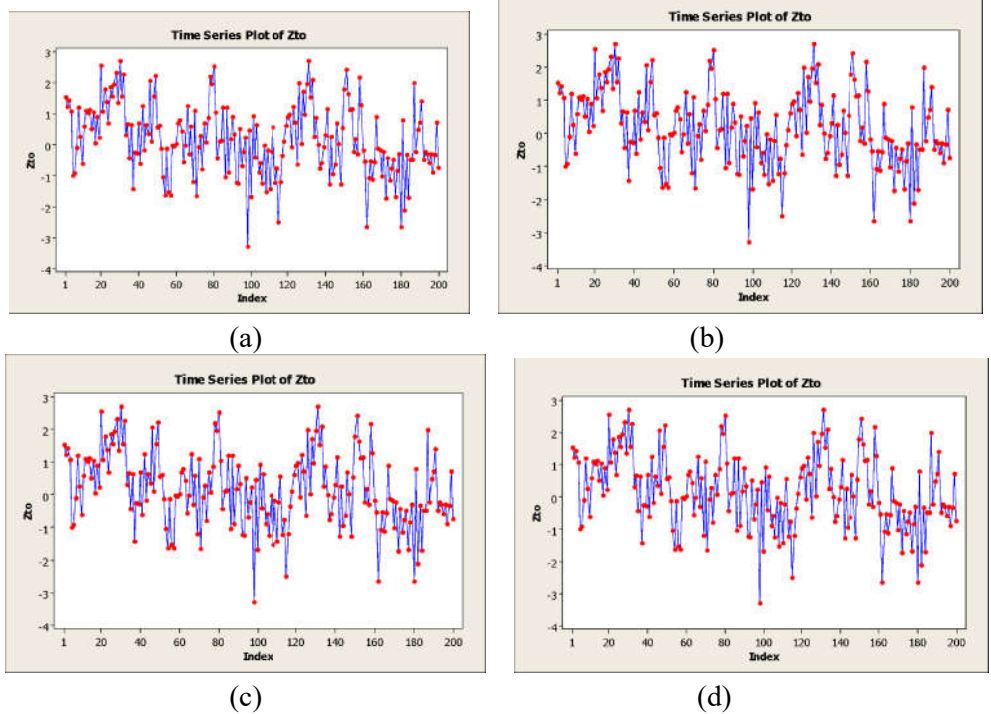

(a)   (b)

(c)   (d)

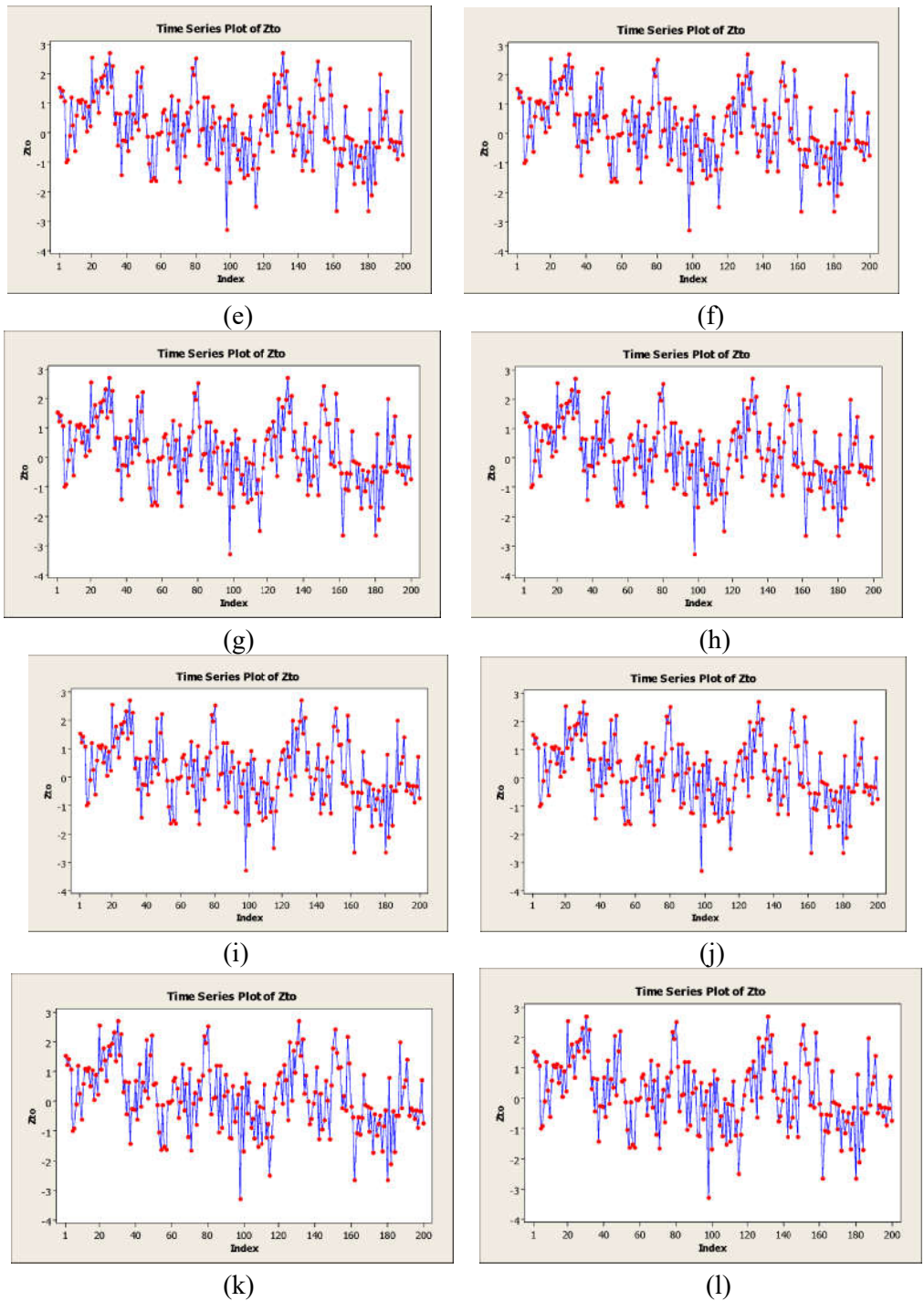

**Figure 2.** (a) normal distribution for the remaining test (Kolmogorov-Smirnov test); (b) Plot Residual Model AR(1); (c) Boxplot Data Residual AR(1); (d) Plot Time Series residues Addition Process Detection Outliers 1$^{th}$; (e) Test the normality of Process Residues Outlier Detection Addition 1$^{th}$; (f) BoxPlot Residue from Process Enhancements Outlier Detection 1$^{th}$; (g) Plot Time Series Process Residues of Outlier Detection Addition 2$^{nd}$; (h) Test the normality of Process Residues the addition of outlier detection 2$^{nd}$; (i) residues BoxPlot outlier detection process Addition 2$^{nd}$; (j) Plot Time Series Process Residues of outlier detection Addition 3$^{rd}$; (k) Testing normality of Process Residues outlier detection Addition 3$^{rd}$; (l) BoxPlot residues The addition of three outlier detection process.

Based on the Kolmogorov-Smirnov test (Figure 2a) obtained p-value <0.010 which is smaller than α=0.05, this shows that the rest do not meet the normal distribution assumption. This is presumably due to the presence of outliers in the data. From the results of the interim model identification, the model AR(2) written mathematically as follows:

$$Z_t = 0,2237 Z_{t-1} + 0,4282 Z_{t-2} + a_t \qquad (24)$$

with MSE of 0.947.

To detect the presence of outliers or outliers in the data is carried out against the residual plot and boxplot of the earlier models we derive the AR(2) model.

From the figure boxplot above shows that there is an outlier at t = 98; thus obtained:

$$I_t^{(T)} = \begin{cases} 1, t = 98, \\ 0, \text{others} \end{cases} \qquad (25)$$

By using the method of the least square regression equation and implementation:

$$Z_t = 0,0600 + 0,209 Z_{t-1} + 0,417 Z_{t-2} \qquad (26)$$

with MSE = 36.780. Then with the addition of outlier detection ARIMA models will be:

$$Z_t = 0,0775 + 0,212 Z_{t-1} + 0,411 Z_{t-2} - 326 X_1(t) \qquad (27)$$

with MSE = 28.194. This step is done continuously in order to obtain a smaller MSE, normality tests are met, and there is no longer an image boxplot outlier data. Here are the results of outlier detection iteration as indicated by improved its MSE.

The addition of a second outlier, namely:

$$I_t^{(T)} = \begin{cases} 1, t = 162, \\ 0, \text{others} \end{cases} \qquad (28)$$

with the regression obtainable:

$$Z_t = 0,0922 + 0,206 Z_{t-1} + 0,410 Z_{t-2} - 327 X_1(t) + 114 X_2(t) \qquad (29)$$

with MSE = 22.808.

The addition of a third outlier, namely:

$$I_t^{(T)} = \begin{cases} 1, t = 180, \\ 0, \text{others} \end{cases} \qquad (30)$$

with the regression obtainable:

$$Z_t = 0,106 + 0,204 Z_{t-1} + 0,401 Z_{t-2} - 329 X_1(t) + 115 X_2(t) + 35,9 X_3(t) \qquad (31)$$

with MSE = 19.365.

With the fulfillment of the residual normality test and its boxplot plot the model fit to the data is ARIMA(2,0,0) with the coefficients obtained from the regression. One of the usefulness of this detection method is that it can reduce the error rate of data prediction. This can be seen in the decline in the value of MSE of each model occurs in the first iteration to the second iteration. The following table of comparisons of each detection and correction and time series models diagram MSE impairment.

**Table 1** Comparison of Model Results

| MODEL | MSE |
|---|---|
| Regression ARIMA (2,0,0) | 36.780 |
| ARIMA (2,0,0) + outlier detection and correction 1th with value: $\omega_{AT} = 0,010215$. $Z_t = 0,0775 + 0,212 Z_{t-1} + 0,411 Z_{t-2} - 326 X_1(t)$ | 28.194 |
| ARIMA (2,0,0) + outlier detection and correction 2nd with value: $\omega_{AT} = -0,0227$ $Z_t = 0,0922 + 0,206 Z_{t-1} + 0,410 Z_{t-2} - 327 X_1(t) + 144 X_2(t)$ | 22.808 |
| ARIMA (2,0,0) + outlier detection and correction 3rd with value: $\omega_{AT} = -0,06622$ $Z_t = 0,106 + 0,204 Z_{t-1} + 0,401 Z_{t-2} - 329 X_1(t) + 115 X_2(t) + 35,9 X_3(t)$ | 19.365 |

## 5. Conclusion

From the discussion and simulation data on a data outlier detection and correction using iterative methods on the ARIMA model (p, d, q) can be obtained several conclusions, as follows:

(1) The process of detection and correction of data containing Additive Outlier on ARIMA (p, d, q) using iteration method and using the software Minitab 16 and Microsoft Excel 2007. The general model for equality Additive Outlier detection ARIMA (p, d, q):

$$Z_t = \sum_{j=1}^{k} \omega_j I_t^{(T_j)} + \pi(B) a_t$$

with: $I_t^{(T)} = \begin{cases} 1, t = T \\ 0, t \neq T \end{cases}$ and $\pi(B) = \dfrac{\theta(B)}{\phi(B)}$

(2) From the simulation results obtained an early model for the data containing outliers i.e., Model AR(2) by the equation: $Z_t = 0,0600 + 0,209 Z_{t-1} + 0,417 Z_{t-2}$ by MSE of 36.780. After the process of outlier detection and correction of data obtained using an iterative method is the best model forecasting model AR (2) with a coefficient obtained from regression and end of the equation:
$Z_t = 0,106 + 0,204 Z_{t-1} + 0,401 Z_{t-2} - 329 X_1(t) + 115 X_2(t) + 35,9 X_3(t)$ by MSE of 19.365.

(3) By using the detection and correction of data containing Additive Outlier on ARIMA (p,d,q) using iterative methods provide process improvements and a decrease MSE models and will certainly provide residual tends to qualify the normal distribution. In this simulation shows that there is improvement MSE value of 47.34% of the initial model.

## References

[1] Ahmar, A. S., Rahman, A., Arifin, A. N. M., & Ahmar, A. A. 2017. Predicting movement of stock of "Y" using sutte indicator. *Cogent Eco. Finance,* **5**, 1 doi:10.1080/23322039.2017.1347123

[2] Rahman, A., & Ahmar, A. S. 2017. Forecasting of primary energy consumption data in the united states: A comparison between ARIMA and holter-winters models. Paper presented at the *AIP Conference Proceedings,* **1885**, doi:10.1063/1.5002357

[3] Wei, W. W. S. 2006. *Time Series Analysis : Univariate and Multivariate Methods* (2nd ed.). New York, NY: Pearson Addison Wesley.

[4] Box, G. E. P., & Cox, D. R. 1964. An analysis of transformations. *J. of Royal Stat. Soc. Series B (Method.)*, 211–252.

[5] Box, G. E. P., & Jenkins, G. M. 1976. Time series analysis, control, and forecasting. *San Francisco, CA: Holden Day*, **3226**(3228), 10.

[6] Yule, G. U. 1926. Why do we sometimes get nonsense-correlations between Time-Series: a study in sampling and the nature of time-series. *J. of Royal Stat. Soc.*, **89**(1), 1–63.

[7] Walker, G. 1931. On periodicity in series of related terms. Proceedings of the Royal Society of London. Series A, Containing Papers of a Mathematical and Physical Character, **131**(818), 518–532.

[8] Slutzky, E. 1937. The summation of random causes as the source of cyclic processes. *Econometrica: J. of Eco. Soc.*, 105–146.

[9] Wold, H. 1938. *A Study in the Analisys of Stationery: Time Series*. Almqvist & Wiksells Boktryckeri.

[10] Makridakis, S., Wheelwright, S. C., & Hyndman, R. J. 2008. *Forecasting methods and applications*. John Wiley & Sons.

[11] Wei, W.W.S., 1994 Time series Analysis:Univariate and Multivariate Methods, Addison-Wesley Publishing Company, California.

[12] Gounder, M.K., Mahendran, dan Rahmatullah. 2007. Detection of Outlier in Non-linear Time series: A Review. Festschrift in honor of Distinguished Professor Mir Masoom Ali. pages 213-224